\documentclass{iaus}
\usepackage{graphicx}

\title[Polaris: Mass and Multiplicity] %% give here short title %%
{Polaris: Mass and Multiplicity}

\author[Evans, et al.]   %% give here short author list %%
{Nancy Remage Evans$^1$, Gail Schaefer$^2$, Howard E. Bond$^2$, 
Edmund Nelan$^2$,
Giuseppe Bono$^3$, Margarita Karovska$^1$, Scott Wolk$^1$,  
Dimitar Sasselov$^4$, Edward Guinan$^5$, Scott Engle$^5$, 
Eric Schlegel$^6$, and Brian Mason$^7$}

\affiliation{$^1$SAO, 60 Garden St., Cambridge MA 02138, USA \break: email 
nevans@cfa.harvard.edu\\[\affilskip]
$^2$STScI, 3700 San Martin Dr., Baltimore, MD 21218, USA\\[\affilskip]
$^3$Univ. Roma, Rome, Italy\\[\affilskip]
$^4$Harvard University, 60 Garden St., Cambridge MA 02138, USA\\[\affilskip]
$^5$Villanova Univ., Dept. of Astronomy, Villanova, PA 19085 USA\\[\affilskip] 
$^6$Univ. Texas, San Antonio, Dept. of Physics and Astronomy, 6900
N. Loop 1604 West, San Antonio TX 78249-0697 USA\\[\affilskip]
$^7$US Naval Observatory, 3450 Massachusetts Ave., NW,
Washington, D.C. 20392-5420, USA}

\pubyear{2007}
\volume{240}  %% insert here IAU Symposium No.
\pagerange{}
\date{?? and in revised form ??}
\jname{Proceedings Title IAU Symposium}
\editors{W. I. Hartkopf, P. Harmanec \& E. Guinan, eds.}
\begin{document}

\maketitle

\begin{abstract}
Polaris, the nearest and brightest  classical Cepheid, is a member of
at least a triple system.  It has a wide ($18''$) physical companion, the 
F-type dwarf Polaris B. Polaris itself is a single-lined  spectroscopic  
binary with an orbital period of 30 years
(Kamper, 1996, JRASC, 90, 140). By combining {\it Hipparcos} measurements of
the instantaneous proper motion with long-term measurements and the Kamper 
radial-velocity  orbit,  Wielen et al. (2000, A\&A, 360, 399) have  predicted
the  astrometric  orbit of the close  companion.  Using the {\it Hubble  Space
Telescope} and the Advanced Camera for Surveys' High-Resolution Channel with
an ultraviolet (F220W) filter, we have now directly detected the close 
companion.  Based on the Wielen et al. orbit, the {\it Hipparcos} parallax, 
and our measurement of the separation ($0.176''$ $\pm$ $0.002''$),
we find a preliminary mass of
5.0 $\pm$ 1.5 M$_{\odot}$ for the Cepheid and 1.38 $\pm$ 0.61 M$_{\odot}$ 
for the close companion. 
These  values will be refined by additional      
{\it HST} observations scheduled for the next 3 years.

We have also  obtained a {\it Chandra}  ACIS-I image of the Polaris field. 
Two distant companions C and D are not X-rays sources  and hence are not
young  enough  to be  physical  companions  of  the  Cepheid.  There  is one
additional stellar X-ray source in the field, located $253''$ from Polaris A, 
which is a possible companion. Further investigation of such a distant 
companion is valuable to confirm the full extent of the system.

\end{abstract}

\firstsection % if your document starts with a section,
\section{Introduction}

Polaris, like most massive stars, is a member of a multiple system.  It is also a 
supergiant (F5 Ib) which is the nearest and brightest classical 
Cepheid.  It is a low amplitude somewhat quirky Cepheid
with a variable amplitude, which is 
pulsating in the first overtone mode, as shown by the {\it Hipparcos}
parallax (Feast and Catchpole, 1997)

The goals of this discussion are to present preliminary results on 
two topics. The first is a direct dynamical measurement of the mass, which 
is of particular interest because it is a Cepheid.  The second is to 
explore how many physical companions belong to the system,
which is of interest as the ``footprints" of star formation.  

\section{Mass}

In order to determine the mass, an orbit must  be derived.  Polaris is a
member of a 30-year spectroscopic system.  We have used the orbit redetermined by
Kamper (1996).  He includes high precision velocities, which 
is important since the orbital velocity amplitude is only 4 km sec$^{-1}$.  
Wielen et al
(2000) made an important breakthrough when they used {\it Hipparcos} proper 
motions to derive an inclination for the system.  

However, no further information could be determined about the mass because 
the companion had never been detected, nor the separation measured.  
We have obtained a the {\it Hubble Space Telescope} ({\it HST}) 
Advanced Camera for 
Surveys (ACS) High Resolution Channel (HRC) image on Aug. 2, 2005 (Fig. 1).
  The companion can be seen in Fig. 1 (left) at about 
7 o'clock.

\begin{figure}
 \includegraphics{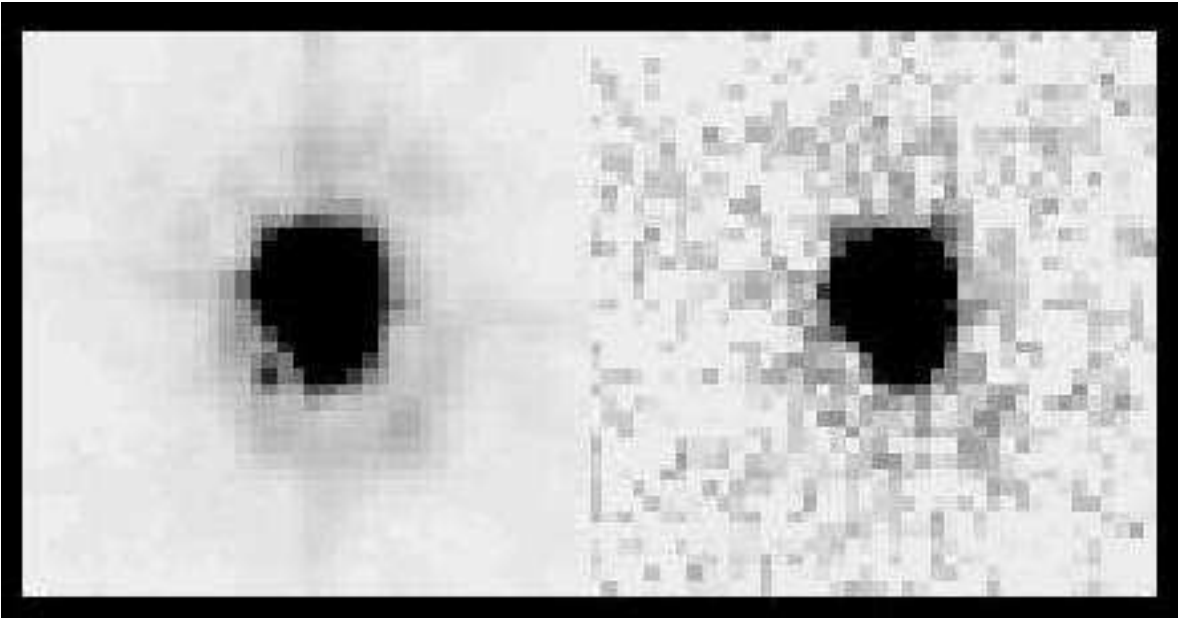}
  \caption{Left: Polaris Aa + Ab taken with the {\it HST } ACS HRC F220W filter
  (coadded images).
  The companion Ab can be seen at approximately 7 o'clock.  The image is 
  approximately $1''$ on a side.  
  Right: Polaris B from the same image  shown to the same stretch.
  }\label{fig:wave}
\end{figure}

Because the point-spread function (PSF) is not completely symmetric, 
and faint wings can be seen in Fig. 1, we have performed several 
tests to confirm the companion detection.  The companion Polaris B is 
also in the images, and is  shown in Fig. 1 (right).  It is noisier 
than A since
B is 6 mag fainter.  However the PSF is very similar, and there
is no artifact at the location of the companion of Polaris A = Polaris Ab.  
In addition, 
we have  examined several images of single white dwarfs taken with the 
same instrumentation from the {\it  HST} archive, and also find 
no indication of an artifact at the location of Ab.  
For a full discussion of the 2005 observation and a second image 
obtained in 2006, see Evans, et al. (2007).

Combining the orbit with the separation measured from the {\it HST} 
images  ($0.176''$ $\pm$ $0.002''$),            
 we find a mass of                                                   
 5.0 $\pm$ 1.5 M$_{\odot}$ for the Cepheid and 1.38 $\pm$ 0.61 M$_{\odot}$
 for the close companion.
 These values are preliminary and will be refined through successive 
{\it  HST} observations.

\section{Companions}  

The second aspect of this study is an investigation of the number of 
members of the system.  While there are many ways to detect a 
binary companion in a system, it is much more difficult to
be certain that the full list of members has been identified.

In the Polaris system, there are 2 highly probable companions, 
Ab, the member of the spectroscopic system, and Polaris B.
Polaris B  ($19''$
distant) is  a probable physical
 companion on velocity grounds (Kamper 1996).  
Two fainter, more distant stars (C and D) might be companions and  
would be K stars if they are at the distance of Polaris.  

Does a system with three stars more massive than the sun have a number of 
low-mass companions as would be predicted by the initial mass function? 
In order to investigate this, we used the following approach.  Any 
low-mass companions (mid-F spectral type and later) 
as young as the Cepheid would produce X-rays.  See, for instance, the study of 
the $\alpha$ Per cluster (Randich, et al. 1996)

We have obtained a 10 ksec  {\it Chandra} image to look for 
low-mass companions.  Stars C and D do 
not appear on the X-ray image, and hence are not
young companions.  Their motion has also been found (by BDM) to be 
incompatible with the Polaris system.  We do find X-rays at the location of Aa + Ab and are working to determine whether they come 
from the Cepheid or the companion.  B was not detected in X-rays, but that
is not surprising for an early F star.  In addition there are a number 
of background AGNs in the image.  They can be distinguished from stars 
since they do not have optical counterparts on 
the 2MASS images, whereas stars at the distance of Polaris 
do have counterparts.  
However, there is one X-ray source which does have a 2MASS counterpart
consistent with an early M star.  This source ``E" becomes the 
one candidate for a further member of the system in the $16'$ $\times$ $16'$
{\it Chandra} ACIS-I field.  

    In summary, only a late M star would have remained undetected on the 
{\it Chandra} image.  This means we have searched for companions to 
approximately a mass ratio of 0.1.  The full content of the 
system is made up of the Cepheid and two 
probable companions with one possible additional low-mass companion.

\begin{acknowledgments}
Support for this work was provided by grants HST-GO-10593.01-A and NAS8-03060,
and also  Chandra  grant  GO6-7011A  and Chandra  X-ray  Center NASA  Contract
NAS8-39073

\end{acknowledgments}

\end{document}